\documentclass[prc,twocolumn,showpacs,twoside,showpacs]{revtex4}

\usepackage{graphicx}
\usepackage{dcolumn}
\usepackage{bm}
\usepackage{CJK}

\begin{document}
%\begin{CJK*}{GBK}{song}

\title{Particle-vibration coupling effect on the $\beta$-decay of magic nuclei}
\author{Y. F. Niu   $^{1,2}$ }
\email{nyfster@gmail.com}
\author{Z. M. Niu  $^3$}
\email{zmniu@ahu.edu.cn}
\author{G. Col\`{o} $^{4,1}$}
\email{gianluca.colo@mi.infn.it}
\author{E. Vigezzi $^{1}$}
\email{vigezzi@mi.infn.it}

\affiliation{$^1$ INFN, Sezione di Milano, via Celoria 16,
I-20133 Milano, Italy}
\affiliation{$^{2}$ Institute of Fluid Physics, China Academy of
Engineering Physics, Mianyang 621900, China}
\affiliation{$^{3}$ School of Physics and Material Science, Anhui University,
             Hefei 230601, China}
\affiliation{$^4$ Dipartimento di Fisica, Unversit\`{a} degli Studi
di Milano, via Celoria 16, I-20133 Milano, Italy}

\date{\today}

\begin{abstract}
Nuclear $\beta$-decay in magic nuclei is investigated, taking into account the coupling between particle and collective vibrations,
on top of self-consistent random phase approximation calculations based on Skyrme density functionals.
The  low-lying Gamow-Teller strength is shifted downwards and at times becomes fragmented;
as a consequence, the $\beta$-decay half-lives are reduced due to
the increase of the phase space available for the decay.
In some cases, this leads to a very good agreement between theoretical and experimental lifetimes:
this happens, in particular, in the case of the Skyrme force SkM*, that can also
reproduce the line shape of the
high energy Gamow-Teller resonance as it was previously shown.
\end{abstract}
 \maketitle
\date{today}
%%%%%%%%%%%%%%%%%%%%%%%%%%%%%%%%%%%%%%%

Weak interaction processes involving atomic nuclei have been the object
of continuous interest for many decades \cite{Grotz:1990}.
The simplest process, that is, nuclear $\beta$-decay,
is mainly determined by the allowed Gamow-Teller (GT) type of transition \cite{Osterfeld1992,Ichimura:2006,Fujita:2011}.
As a rule, most of the GT strength is concentrated
in the high-energy GT resonance (GTR); consequently, the fraction
that contributes to the $\beta$-decay, being in the energetically
allowed region called ``$\beta$-decay window'', is small.
The distribution of the GT strength
is governed both by the nuclear shell structure (in particular,
by the spin-orbit splitting between neutron and proton states), and
by the still underconstrained spin-isospin
channel of the nuclear effective interaction, so that
$\beta$-decay and GTR measurements can complement each other
and be of great usefulness.

In nuclear astrophysics, the $\beta$-decay half-lives set the timescale
of the rapid neutron capture process ($r$-process), and hence influence
the production of heavy elements beyond iron in the universe
\cite{Burbidge1957, Langanke2003, Qian2007}. Other weak processes like
electron capture are of paramount importance in the simulation of
the latest stages of the evolution of massive stars and in the
understanding of possible supernova explosions.
In particle physics, nuclear $\beta$-decay was used to obtain the first
experimental evidence of parity-violation. The
superallowed Fermi $\beta$-decay of nuclei can be exploited to verify the
unitarity of the Cabibbo-Kobayashi-Maskawa
matrix \cite{Towner2010,Liang2009}.

Extensive studies of  $\beta$-decay have been carried
out both experimentally and theoretically. In experiment, important
advances in the measurement of the decay half-lives have been achieved
in recent years with the development of radioactive ion-beam facilities.
As for theory, many approaches have been formulated
ranging from gross theory \cite{Takahashi1973} to the interacting boson-fermion model
\cite{Dellagiacoma1989}.
{\em Ab-initio} approaches can still be used only for light nuclei \cite{Barrett:2013}, whereas the nuclear shell-model can only cover the nuclear chart up to
intermediate values of A $\approx$ 40-50 and/or around magic
regions if some frozen core is assumed.
It performs quite well
in reproducing $\beta$-decay half-lives in these cases
(cf., e.g., the extensive studies in the $sd$ shell \cite{bw85}, in the
$pf$ shell \cite{Langanke2001ADNDT}, and in heavy nuclei \cite{Martnez-Pinedo1999,Suzuki2012}; see also \cite{Li:2014} and references therein).

Another approach, which  can be applied throughout the whole isotope chart, is based on the random phase approximation (RPA), and on its extension to superfluid systems, namely the quasi-particle RPA (QRPA).
Many versions of QRPA  have been applied for
the study of $\beta$-decay, and most of them are based at least
in part on phenomenological ingredients
\cite{Borzov:2006,Fang:2013,Homma:1996,Moller1997}.
Discussing the pros and cons of the various phenomenological
inputs is outside the scope of this work. Certainly, it would be
desirable to reproduce the $\beta$-decays half-lives within
the framework of a self-consistent model without adjustable parameters.

Actually, nuclear $\beta$-decay has also been investigated
 within self-consistent (Q)RPA models based either on
nonrelativistic or relativistic effective interactions (or energy
functionals); in these approaches the
$\beta$-decay half-lives are usually overestimated
\cite{Engel1999,Niksic2005,Marketin2007,Niu2013}. This deficiency can be to a good extent
cured in the case of open-shell nuclei with the inclusion of
an attractive isoscalar proton-neutron (pn) pairing force, that can shift
some GT strength downwards and thus reduce the half-lives. However,
the isoscalar pn pairing has no, or little, effect on closed-
or subclosed-shell nuclei like for example $^{78}$Ni and $^{132}$Sn
\cite{Engel1999,Niksic2005,Marketin2007,Niu2013}.
Thus, the half-lives of such nuclei cannot be reduced; in some cases
$^{132}$Sn is even predicted to be stable \cite{Niksic2005}. Therefore, there
must exist other correlations, other than isoscalar pn pairing, that
are capable to reduce such half-lives. One possible candidate is an
attractive tensor force, as it has been pointed out in Ref.
\cite{Minato2013}. It must be stressed that the introduction of
new terms in the Hamiltonian, like isoscalar pn pairing or tensor
terms, requires the tuning of one or more parameters. It is of
considerable interest to check whether the inclusion of new
correlations {\em without the fitting of new parameters} can
improve the agreement of $\beta$-decay half-lives with experiment.
A recent study \cite{Severyukhin2014} suggests that coupling of the GT states with two-phonon states, as well as the effect of the tensor force, increases the $\beta$-decay rates. However, the model used in  Ref. \cite{Severyukhin2014} is not fully self-consistent since it adopts the Landau-Migdal approximation for the residual force, and makes a strong selection of two-phonon states.
To investigate the role of correlations beyond mean-field in a self-consistent model and in a larger model space,
we apply for the first time the RPA plus particle-vibration coupling (PVC) to
the study of $\beta$-decay.

The RPA approach for giant resonances
or other vibrational
states is restricted to configurations of one particle-one hole
(1p-1h) nature.  In order to reproduce the observed spreading
widths, one must consider the damping  caused by the
coupling to more complicated states, like
2p-2h configurations \cite{Kuzmin1984,Drozdz:1990,Dang:1997}.
An effective way
to account for (most of) the observed spreading widths is to include
the contribution of 1p-1h-1 phonon configurations or, in other
words, to take into account the
coupling of single-nucleon states to the collective low-lying (mainly
surface) nuclear vibrations or phonons  \cite{Bertsch1983}.
We call this model
RPA plus PVC, and we note that self-consistent versions of such a model
have been realized based on relativistic and nonrelativistic
interactions or energy functionals.  Good
agreement with the experiment  has been obtained
in the case of line shape of the
GT and spin-dipole strength distribution
\cite{Litvinova2014,Marketin2012,Niu2012,Niu2014}. In our application of
the self-consistent RPA+PVC model, based on Skyrme energy density functionals
\cite{Niu2012,Niu2014}, it was found  that the coupling to phonons
produces a downward shift of the GT excitation strength distribution (by about 1-2 MeV),
accompanied by the development of broadening and/or fragmentation.
In short, we expect that correlations beyond 1p-1h ones, and
due to PVC, can play an important role also in the $\beta$-decay
processes. Accordingly, in this Letter,
we apply our Skyrme RPA plus PVC model
to the study of $\beta$-decay. Our model is self-consistent in the sense
that the same Skyrme force is used to calculate the single-particle
levels and the RPA spectrum (both the GT one and those of
the low-lying surface vibrations to be coupled to it), as well as the
PVC vertices.

The PVC effects are introduced along the line of
Refs. \cite{Niu2012,Niu2014} as we shall now discuss.
The RPA in its standard matrix form produces a set of  $1^+$ eigenstates  $|n\rangle$, having
energies $E_n$ and strengths $B_n$, as well as the forward-going and backward-going amplitudes
denoted by $X^{(n)}_{ph}$ and $Y^{(n)}_{ph}$, respectively.
We then couple these RPA states with a set of doorway states consisting of a p-h excitation coupled
to a collective vibration. These collective vibrations are obtained by computing
the RPA response with the same Skyrme interaction. We have considered phonons with natural
parity $J^{\pi} = 0^+$, $1^-$, $2^+$, $3^-$, $4^+$, $5^-$, and $6^+$ having energies smaller than 20 MeV and  associated with a fraction of the total (isoscalar or isovector) strength larger than $5\%$.

The self-energy of the RPA state $\vert n\rangle$ is given by
\begin{eqnarray}
   \Sigma_n(E_M) &=& \sum_{ph,p'h'} W^\downarrow _{ph,p'h'} (E_M) X_{ph}^{(n)} X^{(n)}_{p'h'} \nonumber\\
  && + W^{\downarrow*} _{ph,p'h'} (-E_M) Y_{ph}^{(n)} Y^{(n)}_{p'h'},
\end{eqnarray}
where the matrix elements $W^\downarrow _{ph,p'h'}(E_M)$ are spreading terms associated with the
coupling of the 1p-1h configurations with the doorway states, defined
in Refs. \cite{Niu2012,Niu2014}.
They are complex and energy-dependent, calculated by using an averaging
parameter $\eta$ that avoids divergences
and represents, in an approximate way,
the coupling of the doorway states to even more complicated
configurations.
We have found that the lifetimes converge for small values of $\eta$,
which are expected to be appropriate for  low-lying, discrete states.  In our calculations we have set
$\eta= 0.05$ MeV.
One can then calculate
the GT strength distribution from the Gaussian averaging \cite{Mahaux1985}
\begin{equation}
  S(E_{M}) =  \sum_n  \frac{1}{\sigma_n \sqrt{2\pi}}
  {\rm e}^ {-\frac{(E_{M}-E_{n} - \Delta E_{n} )^2 }{2\sigma_n^2}} B_n,
\end{equation}
where
$\sigma_n$ is defined as $\sigma_n = (\frac{\Gamma_n}{2}+\eta)/\sqrt{2{\rm ln} 2}$, with
$\Delta E_{n}= {\rm Re } \Sigma_n(E_M) $ and $\Gamma_n=-2 {\rm Im} \Sigma_n(E_M)$.

\begin{figure}
\includegraphics[scale=0.3,angle=0]{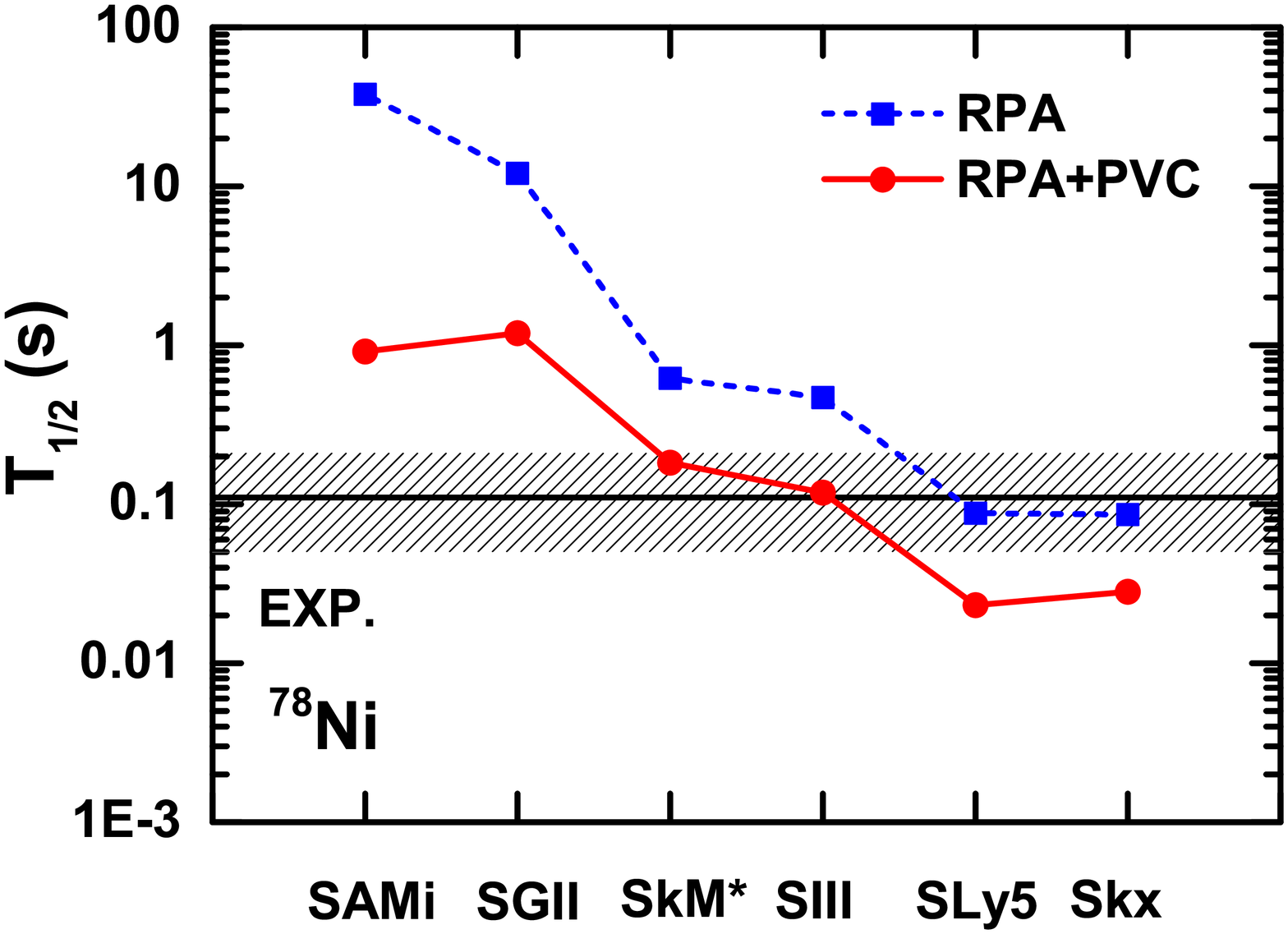}
\caption{(Color online) $\beta$-decay half-life of $^{78}$Ni, calculated by RPA and RPA+PVC
approaches with several different Skyrme interactions, in comparison with the experimental
value~\cite{Hosmer2005}.} \label{fig1}
\end{figure}

Once the strength function has been obtained,
the $\beta$-decay half-life of an even-even nucleus is calculated in the
allowed GT approximation by using the expression
\cite{DeShalit:1974,Engel1999,Niu2013}
\begin{equation}
\label{halflife}
 T_{1/2} =
   \frac{D}{g_A^2 \int^{\Delta_{nH}} S(E_M) f(Z,\omega) dE_M }
\end{equation}
with $D=6163.4$ s and $g_A=1$. The
 integration includes all final $1^+$ states having an
excitation energy $E_M$, referred to the ground state of  the mother nucleus, smaller than  $\Delta_{nH}=0.782$ MeV, which is the
mass difference between neutron and hydrogen.
If the energy is instead referred to the ground state of the daughter nucleus, and is denoted by $E$, one has
$E = E_M - \Delta B$, $\Delta B$ being the experimental binding energy difference $B_M-B_D$.
This choice  is often convenient, because the calculated energy of the final $1^+$ states can  then be directly compared to the experimental spectrum of the final nucleus.
It may happen that  the calculated energy of the lowest state lies at negative energy,  that is, below the experimental value of the ground state
energy of the daughter nucleus.
The upper limit of integration
in (\ref{halflife})
becomes equal to the $Q_{\beta}$ value ($\Delta_{nH} - \Delta B  = Q_{\beta}$),
namely to the atomic mass difference between the mother and daughter nucleus. Eq. (\ref{halflife}) then becomes
\begin{equation}
\label{halflifed}
 T_{1/2} =
 \frac{D}{g_A^2 \int^{Q_\beta} S(E) f(Z,\omega) dE }.
\end{equation}
The integrated phase volume $f(Z,\omega)$ is given by
 \begin{equation}
  f(Z,\omega)= \int_{m_e c^2} ^{\omega}
  p_e E_e (\omega - E_e)^2 F_0(Z+1,E_e) d E_e,
 \end{equation}
where $p_e, E_e$, and $F_0(Z+1,E_e)$ denote the momentum, energy and Fermi
function of the emitted electron, respectively.
$\omega$ is the energy difference between the initial and final nuclear state,
connected with GT energy $E$ (or $E_M$) by $\omega=Q_\beta + m_e c^2 - E=\Delta_{np}-E_M$ where $\Delta_{np}=1.293$ MeV
is the mass difference between neutron and proton.

Fig. \ref{fig1} shows the results for the $\beta$-decay half-life of $^{78}$Ni, a doubly magic nucleus
that is an important waiting point in the $r$-process, and represents as a major bottleneck in the synthesis
of heavier elements.
As recalled in the beginning of this Letter and evident in the figure,
the different Skyrme interactions
are not well constrained in the spin-isospin channel and their predictions for the half-life at
the RPA level can vary over more than two orders of magnitude. For instance, SAMi \cite{Roca-Maza2012} predicts a rather collective GTR and does not leave much strength at low energy, resulting in a long half-life, at
variance with SLy5 \cite{Chabanat1998} that has weak spin-isospin terms and produce a less collective GTR, leaving more
strength at low energy and providing a much shorter half-life. The parameter sets
SAMi, SGII \cite{Giai1981}, SkM* \cite{Bartel1982} and SIII \cite{Beiner1975} overestimate the half-life, while the interactions SLy5 and Skx \cite{Brown1998}
are in agreement with the experiment at the RPA level.

The inclusion of PVC effects reduces the half-lives for all interactions systematically.
The reduction factor  $R$ is larger
for SAMi ($R \approx 42$)  and SGII ($R \approx 10$) while it is equal to about $4$
for the other four interactions.
Within RPA+PVC,
the half-life obtained with the sets SkM* and SIII falls within the experimental error.
It has to be stressed that the Skyrme force SkM* does not only reproduce well
this $\beta$-decay half-life, but also describes well the giant resonance line shape in $^{208}$Pb and $^{56}$Ni
at the PVC level \cite{Niu2012,Niu2014}.
We will show results for the $\beta$-decay
of $^{132}$Sn, $^{68,78}$Ni, and $^{34}$Si with all interactions, but we will discuss in
detail the SkM* results.

\begin{figure*}
\includegraphics[scale=0.2,angle=0]{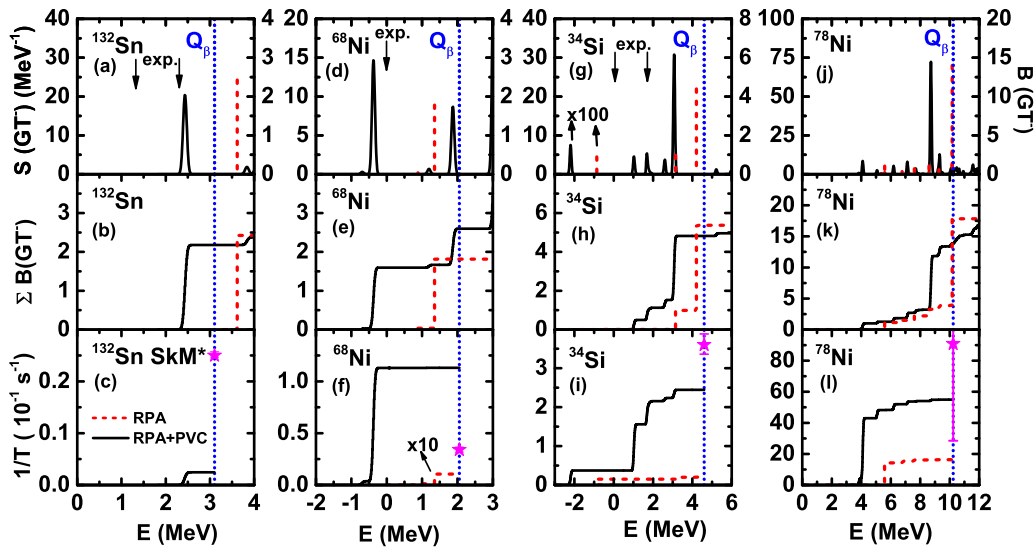}
\caption{(Color online) Experimental data related to $\beta$-decay from  nuclei $^{132}$Sn, $^{68}$Ni, $^{34}$Si,
and $^{78}$Ni are compared with theoretical results obtained with the SkM* interaction.
In these panels, the excitation energies $E_M$ calculated with respect to the
mother nucleus are transformed to $E$, the excitation energies referred to the ground state of daughter nucleus,
using experimental binding energy difference (see the text); accordingly, the vertical dotted lines show the experimental
value of $Q_\beta$ \cite{nndc}.
Top panels: GT$^-$ low-lying strength  associated with the discrete RPA peaks $B$(GT$^-$) (dashed lines)
and with the continuous RPA+PVC strength distributions $S$(GT$^-$) (solid lines).
The arrows indicate the experimental energies of the measured $1^+$ states \cite{nndc}.
Middle panels:
cumulative sum of the RPA and RPA+PVC strength shown in the top panels.
Bottom panels: cumulative sum of $1/T_{1/2}$.
The experimental
values of $1/T_{1/2}$ \cite{nndc} for each nucleus
are indicated by the stars. The strength of the lowest RPA and RPA+PVC peaks in panel (g) and the RPA $1/T_{1/2}$ in panel (f) have been multiplied by a factor of 100 and 10, respectively.} \label{fig2}
\end{figure*}

In order to understand the reasons for the systematic decrease of the half-lives after the
inclusion of phonon coupling, we display in Fig. \ref{fig2} the GT strength distributions
(with respect to the daughter nucleus), the cumulative sums of the strengths, and
the cumulative sums of $1/T_{1/2}$ [that is, the values obtained from Eq. (\ref{halflifed}) when
$Q_\beta$ is replaced by a running $E$ in the upper limit of the integral in the denominator].
Generally speaking, for all nuclei we have considered the GT peaks are shifted downwards when going
from RPA to RPA+PVC.
As seen in Fig. \ref{fig2}(a), in the case of $^{132}$Sn, in RPA there is no strength
below $Q_\beta$, so that the nucleus is stable. The lowest RPA state located at $E=3.6$ MeV
is moved into the $\beta$-decay window and becomes very close
to one of the experimentally observed $1^+$ states (which, however,
has a decay branching ratio of only $I=0.87\%$).
The total strength below $3.6$ MeV then  does not change much from RPA to RPA+PVC [Fig. \ref{fig2}(b)].
Another (lower) $1^+$ state, which experimentally
has by far the largest decay branching ratio,
is not reproduced by our model; thus, our RPA+PVC can predict a finite value of the half-life,
providing a qualitative improvement with respect to RPA but still overestimates the experimental
finding [Fig. \ref{fig2}(c)].
In the case of $^{68}$Ni, RPA predicts a state in the $\beta$-decay window but its energy is
higher than experiment as seen in Fig. \ref{fig2}(d). This explains why the half-life is overestimated
with respect to the experimental finding: the contribution of this state
to $1/T_{1/2}$ is so small that it is only visible in Fig. \ref{fig2}(f) when multiplied by a factor $10$.
With the inclusion of PVC, the RPA peak at $1.5$ MeV is moved even
slightly below the experimental ground state energy.
Although, as in the previous case, the strength of this state is not changed much
as compared with RPA [Fig. \ref{fig2}(e)], it gives a very large contribution to $1/T_{1/2}$ because of the
large phase-space factor. As a consequence, the half-life is smaller than in experiment.
In the case of $^{34}$Si, in RPA one finds three peaks located at $E=-0.86, 3.1$, and $4.2$ MeV.
The first one lies below the experimental
ground state and determines the value of $1/T_{1/2}$ [Fig. \ref{fig2}(i)].
This peak carries a very small value of the strength and therefore   the
experimental lifetime is largely overestimated.
Including the PVC the strength becomes fragmented [Fig. \ref{fig2}(g)].
One can identify five peaks at
$E=-2.2, 1.0, 1.7, 2.6$, and $3.1$ MeV, contributing respectively $15\%, 49\%, 24\%, 3\%$, and $9\%$ of the total
value of $1/T_{1/2}$, which becomes much larger than in RPA,  substantially improving the agreement with the experimental lifetime.
For the nucleus $^{78}$Ni, the small strength at $E=5.6$ MeV gives almost all the contribution
to $1/T_{1/2}$ in the RPA model [Fig. \ref{fig2}(l)] which underestimates the experimental value.
With PVC, the state at $E=5.6$ MeV is shifted to a state peaked at 4.0 MeV [Fig. \ref{fig2}(j)]
with strength unchanged [Fig. \ref{fig2}(k)]. As a result, due to the increase of phase space
its contribution to $1/T_{1/2}$ becomes about $3.4$
times larger than in RPA [Fig. \ref{fig2}(l)].
The strength distribution above this peak contributes $22\%$ of the total $1/T_{1/2}$.

\begin{figure}[!ht]
\includegraphics[scale=0.15,angle=0]{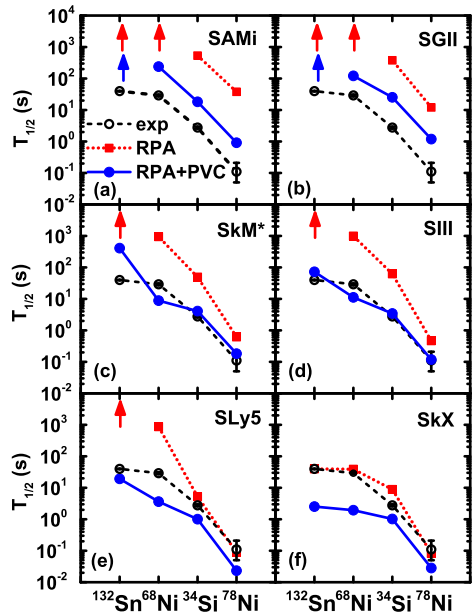}
\caption{(Color online) The $\beta$-decay half-lives of $^{132}$Sn, $^{68}$Ni, $^{34}$Si,
and $^{78}$Ni, calculated by RPA and RPA+PVC approaches, respectively, in comparison with experimental values \cite{nndc}. The arrows denote half-lives longer than $10^6$ s.} \label{fig3}
\end{figure}

The resulting calculated lifetimes for these four nuclei, in the case of all the forces used in Fig. \ref{fig1},
are compared with experiment in Fig. \ref{fig3}.
The RPA results generally markedly overestimate the half-lifes for all nuclei.
An exception is represented by the  interaction Skx, in which case one obtains a  good agreement with data at the RPA level;
this is associated with  the fact that the properties of $^{132}$Sn, $^{68}$Ni, and $^{34}$Si, as well as the single-particle levels of $^{132}$Sn and $^{34}$Si,
have been used to fit of the parameters of this force  \cite{Brown1998}. The effect of the PVC decreases the values of $T_{1/2}$ by
large factors compared to RPA, substantially  improving the agreement
with experimental data, except for Skx and (partially) for SLy5. With the inclusion of PVC effect, the interactions SkM* and SIII give the best
agreement with data. More in detail, in the case of SkM*, the lifetime is still large in $^{132}$Sn and small in $^{68}$Ni, in keeping with the errors in the position of the lowest ${1^+}$  state (cf. Fig. \ref{fig2}).
Theory agrees instead very well with data in the case of  $^{34}$Si and $^{78}$Ni.

In conclusion, we have shown that, starting from RPA, the coupling between particles
and vibrations causes a significant downward shift in the GT strength function of the four nuclei
$^{132}$Sn, $^{68}$Ni, $^{34}$Si, and $^{78}$Ni (treated as magic).
The $\beta$-decay half-life is more sensitive to the position of the 1$^+$
states rather than to the strength, which is not much changed in going from
RPA to RPA+PVC. This is due to the strong increase of the decay phase space factor
as the energy decreases. As a consequence,
the lifetime is reduced in the case of RPA+PVC and
the agreement between theory and experiment is in general substantially improved.
In particular, the interaction SkM* that had been previously shown to perform well
in magic nuclei as far as the line shape of the GT resonance is concerned \cite{Niu2014},
leads to overall good agreement with $\beta$-decay data.

We can expect that including the effect of PVC
will also be helpful in the case of other weak interaction processes. For example,
in the case of the electron capture on fp-shell nuclei, the threshold position
and cross sections at low energies are underestimated with respect to e.g. shell-model
calculations \cite{Fantina2012}. PVC can lower the excitation energies in the
GT$_+$ channel in a similar way as it has been shown here.
The study of open-shell nuclei by including pairing correlations is envisaged.
Then the model can be employed to predict the half-lives of $r$-process bottle neck nuclei with $N=82$, which play an important role for the duration of the r-process, and hence can help to understand the origin of heavy elements in the universe.

Partly supported by the National Natural Science Foundation of China under Grant No. 11305161 and No. 11205004.

\bibliographystyle{apsrev}

%\bibliography{RPAPVC} % Produces the bibliography via BibTeX.

%\end{CJK*}
\end{document}